\documentclass[12pt]{article}
\usepackage{epsfig}

\parskip=\medskipamount 
plus.3em minus.5em \abovedisplayshortskip=.5em plus.2em minus.4em
\renewcommand{\thesection}{\arabic{section}}

\catcode`@=11

\@addtoreset{equation}{section} \@addtoreset{equation}{subsection}
\def\theequation{\ifnum\value{section}=0 \arabic{equation}\ignorespaces
\else \ifnum\value{section}=-1 A.\arabic{equation}\ignorespaces
\else \ifnum\value{subsection}=0
\thesection.\arabic{equation}\ignorespaces \else
\thesection.\arabic{subsection}.\arabic{equation}\ignorespaces
                             \fi
                        \fi
                   \fi}

{\catcode`\'=\active \def'{{}^\bgroup\prim@s}}

\catcode`@=12

\newcommand{\bq}{\begin{equation}}
\newcommand{\be}{\begin{equation}}
\newcommand{\fq}{\end{equation}}
\newcommand{\ee}{\end{equation}}
\newcommand{\bqr}{\begin{eqnarray}}
\newcommand{\beqs}{\begin{eqnarray}}
\newcommand{\fqr}{\end{eqnarray}}
\newcommand{\eeqs}{\end{eqnarray}}

\newcommand{\rf}[1]{(\ref{#1})}

\def\bop#1{\setbox0=\hbox{$#1M$}\mkern1.5mu
    \vbox{\hrule height0pt depth.04\ht0
    \hbox{\vrule width.04\ht0 height.9\ht0 \kern.9\ht0
    \vrule width.04\ht0}\hrule height.04\ht0}\mkern1.5mu}

\begin{document}
\thispagestyle{empty}

\begin{flushright}
\begin{tabular}{l}
hep-th/0503081 \\
\end{tabular}
\end{flushright}

\vskip .6in
\begin{center}

{\bf  A New Knot Invariant}

\vskip .6in

{\bf Gordon Chalmers}
\\[5mm]

{e-mail: gordon@quartz.shango.com}

\vskip .5in minus .2in

{\bf Abstract}

\end{center}

A polynomial is presented that models a topological knot in a
unique manner.  It distinguishes all types of knots including
the orientation and has a group theory interpretation.  The topologies
may be labeled via a number, which upon a base $2$ expansion generate
the polynomial; the equivalent numbers via Reidemeister moves are grouped
into a superset polynomial with coefficients labeling the equivalent
knots.

\setcounter{page}{0}
\newpage

\setcounter{footnote}{0}

\section{Introduction}

The classification of knots topologically has been of interest for
many years, but a unique invariant appears to be lacking in the
literature.  In this paper a unique invariant is given.

There are several forms of knot invariants written in polynomial
form, and they are of both mathematical and physical interest
\cite{EDM2},\cite{Rolfsen}.  An
invariant that distinguishes all topologies from each other is
relevant for many reasons.

The invariant presented here relies on labeling all intersections
of the curve in three dimensions by two by two matrices.  These
two-by-two matrices are assembled into a larger matrix which could
serve as an invariant; however, both for notational purposes and
to make contact with previous forms this larger matrix is projected
onto Sp($2n$) adjoint generators into a polynomial form.

The knot is first labeled in the manner: (1) a starting point is
chosen on the contour, (2) the knot is given a direction by attaching
arrows one way through the contour, (3) a number is attached to
every intersection along this direction post (or prior) to every
intersection, and (4) each intersection of the contour with itself
takes on only one of four forms and is labeled by two numbers
generated in (3).  Furthermore, the four types of oriented
intersections are illustrated in the figure 1(a).

These four types of oriented intersections are labeled with a two
by two matrix. These matrices are,

\bqr
M_1 = \pmatrix{ 1 & 0 \cr 0 & 0 }
  \qquad M_2 = \pmatrix{ 0 & 1 \cr 0 & 0 }
\fqr
\bqr
M_3 = \pmatrix{ 0 & 0 \cr 1 & 0 }
  \qquad M_4 = \pmatrix{ 0 & 0 \cr 0 & 1 } \ .
\label{intersections} \fqr There are a total of $n$ intersections
in the knot configuration, which through a single closed contour
are passed through twice each in traversing the loop.  These
matrices are assembled into a $2n$ by $2n$ matrix $M$ via block
form by inserting at position (i,j) the two by two matrix
associated with the (i,j) node along the contour; this fills up
all but the diagonal elements.  The diagonal entries along (i,i)
are given an empty two by two matrix.  Also, via following the
arrows, the lower triangular two by two matrices are the transpose
of the upper triangular ones and the matrix satisfies $M=M^T$.
(Up and then under to the right, $M_1$, is the transpose of
passing through the intersection along the path of the other
arrow, which is up and then over to the right, $M_4$).
\begin{figure}
\begin{center}
\epsfxsize=12cm
\epsfysize=12cm
\epsfbox{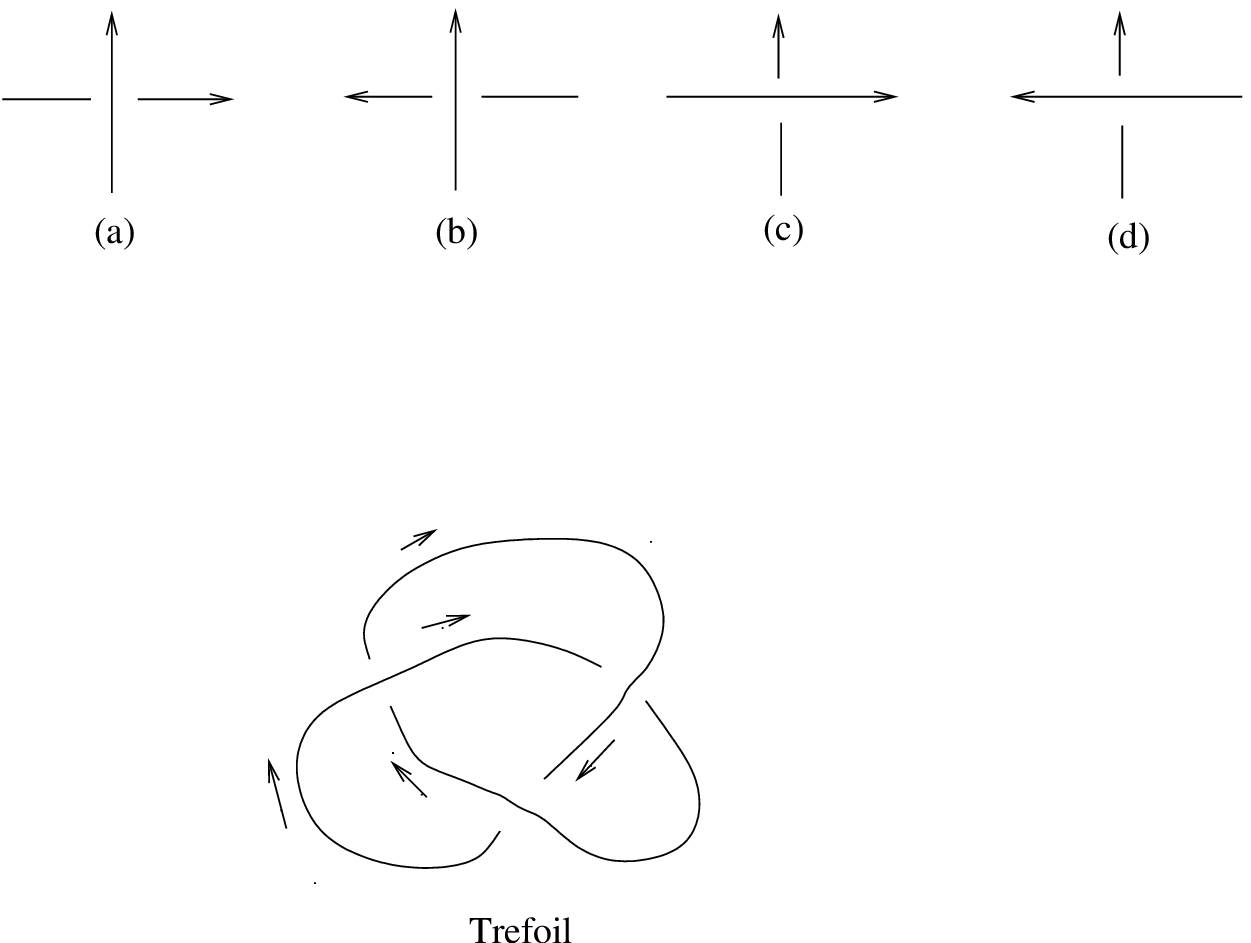}
\end{center}
\caption{(1) The four types of intersections.  (2) A sample
trefoil knot.}
\end{figure}

This matrix is a member of Sp($2n$) and allows a projection onto the
adjoint representation, $M=\sum_i a_i T^i$.  Note that all entries are
unity, which means that the knot matrix is associated with the
homology of a (possibly degenerate) Riemann surface $\Sigma_n$ of
genus $n$.  Without loss of information, one could put minus signs in
the upper triangular portion so that the final matrix satisfies $M=-M^T$,
i.e. a member of SO($2n$).  The Sp($2n$) (or SO($2n$)) generators could
be given the standard form,

\bqr
(M_{ab})^{ij} = \delta_a^i \delta_b^j \pm \delta_a^j \delta_b^i \ .
\label{matrixelement}
\fqr

The polynomial invariant is constructed from the topology of the knot,
in $M$, via the projection $M=\sum a_i T^i$.  The coefficients $a_i$
from this explicit projection are assembled into the form $P(z)$,

\bqr
P(z) = \sum_{i=1}^{2n} a_i z^i \ .
\label{invariant}
\fqr
The invariant in \rf{invariant} is unique and distinguishes all of
the possible topologies, because the matrix uniquely reconstructs the
knot and there is no loss of information between $M$ and $P(z)$.
There is an ambiguity in mapping the coefficients in the matrix
decomposition $M=\sum a_i T^i$ to the polynomial invariant in
\rf{invariant}.  The ambiguity is removed via labeling one to one in
order $T^i \leftrightarrow z^i$.

As an illustration of the procedure, one of the two trefoil knots in
figure 1(b) is analyzed.  The trefoil knot has three intersections and
so is dimension twelve.  The associated matrix $M_t$ written in block
form with the $M_j$ matrices is,

\bqr
M_t = \pmatrix{ 0 & 0 & 0 & 2 & 0 & 0 \cr
                0 & 0 & 0 & 0 & 3 & 0 \cr
                0 & 0 & 0 & 0 & 0 & 2 \cr
                3 & 0 & 0 & 0 & 0 & 0 \cr
                0 & 2 & 0 & 0 & 0 & 0 \cr
                0 & 0 & 3 & 0 & 0 & 0 } \
\fqr
The decomposition of this trefoil's $M_t$ is $a_{8,1}=1$, $a_{9,4}=1$,
and $a_{12,5}=1$ (with symmetrization).  The polynomial $P_t(z)$ is, via
the decomposition of the generators through $z^{(j-1)*2n+i}$,

\bqr
P_t(z) = z^8 + z^{40} + z^{60} \ .
\fqr
Note that this labeling of the generators has vanishing elements for
diagonal elements $i=j$. This simple example describes the procedure for
finding $M$ and $P(z)$.  It is not clear if this polynomial can be given
further number theoretic or geometric interpretation due to the appearance
of the numbers $8*(1,5,12)$.

The Reidemeister moves may also be examined in this context.  There
are three moves that are analyzed.  The first one involves an overlap
with a twist, depicted in figure 1, and amounts to an expansion of the
matrix $M$ in the $i$ row and $i+1$ column with the matrix entry $M_1$,

\bqr
z^{2(i-1)n+2(i+1)+1} \ ,
\fqr
while changing the rest of the matrix with zeros in the $i$th row and
and $i+1$th column, via a $M_1$.  The second Reidemeister move involves
the inclusion of two additional matrices $M_4$ and $M_2$, at the nodes
$i$, $j+1$ and $i+1$, $j$.  This involves enlarging the matrix $M$ by
the terms

\bqr
z^{2in+2(j+1)} + z^{2in+2j} \ ,
\fqr
with zeros placed in the columns and rows of the entries at $i$, $j+1$
and $i+1$, $j$.  The third move involves the triple crossing, i.e. a
slide of a bar, from the entries $M_4$ at $(i,j)$, $M_2$ at $(j+1,k+1)$
and $M_4$ and $(i+1,k)$; to the entries  $M_3$ at $(j,k)$, $M_1$ at
$(k+1,i)$, and $M_1$ at $(j+1,k+1)$.  This involves the change of the
entries from these nodes from,

\bqr
z^{2in+2j} + z^{2jn+2(k+1)} + z^{2(i+1)n+k} \
\fqr
to

\bqr
z^{jn+2(k-1)+1} + z^{kn+2(i-1)+1} + z^{jn+2k+1} \ .
\fqr
These Reidemeister moves may be incorporated directly at the level
of the polynomials $P(z)$ or in the matrices $M$.

The polynomial form of the invariant $P(z)=a_i z^i$ with the unit
coefficients $a_i$ may be given a base $2$ interpretation via the
expansion of a number

\bqr
N = a_i z^i
\label{numberform}
\fqr
with the expansion over the base $2$ numbers $z^0=1$, $z^1=2$, $z^2=4$,
etc.  Not all numbers $N$ may be reached via the expansion due to the
expansion of the matrices $M_1$, $M_2$, $M_3$, and $M_4$.  However,
another interpretation is given in base $4$ via the expansion of the
matrix invariant with the labels $1$ through $4$.  Considering the
equivalence of the knots via the Reidemeister moves, a family
of equivalences may be defined via a new polynomial $Q_N(z)$,

\bqr
Q_N(z) = \sum b_i w^i \ ,
\label{equivalence}
\fqr
with the first coefficient $b_0$ defining the fundamental (minimal)
knot.  The coefficients $b_i$ are numbers labeling further knots
related to the minimal knot via Reidemeister moves.  These numbers
are base two (or base four), spanning the knot topology via the
expansion,

\bqr
b_i = \sum a_j^{i} z^i  \ ,
\fqr
with the $b_i$ essentially $P(z)$.  The tower of numbers $b_i$
may be obtained by direct calculation or an iteration of the fundamental
knot.  There is potentially interesting group theory characteristics,
e.g. representation dimensions, associated with the numbers $b_i$.
For example, the individual equivalence classes form separate fields,
subsets of the integers, which are closed under the Reidemeister moves.

The invariant $P(z)$ is unique and completely characterizes the knot
configuration; multiple disconnected but entangled contours are also
described via the labeling of the intersections.  Due to the construction
this invariant has a group theoretic symplectic interpretation.  The
matrix forms $M$ of the polynomials could be investigated further for
more information (e.g. invariants of matrices, embeddings of one knot
into another, quotients, $\ldots$).  Furthermore, the matrix form has
an interpretation in terms of the homology of a max genus $n$ Riemann
surface.

The polynomial form should have relations to other commonly used invariants
such as the HOMFLY, Jones, Kauffman, or Vassiliev ones.  Although
these latter forms do not uniquely specify the knot configuration,
the relation is relevant to physics models and mathematics.

Because the invariant $P(z)$ is unique, the classification and further
development of associated three-dimensional Seifert manifolds, such as
cohomology directly from $P(z)$, may be found in a more direct fashion.
The algebraic nature of the knot further relates to geometry in $d=2$
via the zero set $P(z)=0$.

Last, the invariants $P(z)$ presented here always have unit coefficients.
The information is encoded in the exponents $i$ in the expansion $P(z) =
\sum a_i z^i$.  Other invariants are typically of lower degree, but with
non-unity in the (seemingly less sparse) coefficients; the $P(z)$ here
contain more information in the exponents apparently.  The informaton
content is the same however, apart from the uniqueness issue.  In
comparison between the coefficients and exponents, it is not obvious how
many bytes of information the different forms require to label a knot.

The equivalence classes of the knot numbers via the Reidemeister moves
is found via the polynomial operations.  These have an indirect number
form $f_\sigma(i)(N)$ for the actions $\sigma(i)$ of the moves $i$ on
the knot number $N$.

\vfill\break

\end{document}